\documentclass[conference]{IEEEtran}

\iffalse
\else
\fi

\usepackage{ucs}
\usepackage{array}
\usepackage[utf8x]{inputenc}
\usepackage{amsmath}
\usepackage[linesnumbered,ruled]{algorithm2e}
\usepackage{algorithmicx}
\usepackage{makecell}
\usepackage{graphicx}
\usepackage{subcaption}
\usepackage{multirow}
\usepackage{layout}
\usepackage{cite}
\usepackage{setspace}
\usepackage{listings}

\usepackage{float}

\floatstyle{ruled}
\newfloat{program}{thp}{lop}
\floatname{program}{Program}
\usepackage[a4paper,bindingoffset=0.2in,%
left=0.6in,right=0.6in,top=0.9in,bottom=0.9in,%
footskip=.2in]{geometry}

\usepackage{color, colortbl}
\definecolor{Gray}{gray}{0.9}
\definecolor{LightCyan}{rgb}{0.88,1,1}

\definecolor{dkgreen}{rgb}{0,0.6,0}
\definecolor{gray}{rgb}{0.5,0.5,0.5}
\definecolor{mauve}{rgb}{0.58,0,0.82}

\usepackage[T1]{fontenc}
\usepackage{array}
\usepackage{makecell}
\newcolumntype{x}[1]{>{\centering\arraybackslash}p{#1}}

\usepackage{tikz}

\newcommand{\OpCosts}{\mathit{OpCosts}}
\usepackage[T1]{fontenc}
\usepackage{mathptmx}

\begin{document}%\rmfamily%\fontsize{13}{13}
	
	\title{A Source-level Energy Optimization Framework for Mobile Applications}

	\author
	{\IEEEauthorblockN{Xueliang Li\IEEEauthorrefmark{1} \quad
		John P. Gallagher\IEEEauthorrefmark{1}\IEEEauthorrefmark{2}	}
		\IEEEauthorblockA{\IEEEauthorrefmark{1}Roskilde University, Roskilde, Denmark}
			%Email: xueliang@ruc.dk}
		
	    \IEEEauthorblockA{\IEEEauthorrefmark{2}IMDEA Software Institute, Madrid, Spain\\ Email: \{xueliang, jpg\}@ruc.dk  }
		%Email: jpg@ruc.dk}
		%\and
		%\IEEEauthorblockN{}
		%\IEEEauthorblockA{Roskilde University\\
		%Email: jpg@ruc.dk}
	}

	
	%Telephone: (800) 555--1212, Fax: (888) 555--1212}


	\maketitle
	
	\begin{abstract}
		%\linespread{0.3} 

		%however,  effective energy optimization relies mainly on developers, since compiler-based energy optimizations are limited. In this paper, we propose an energy-optimization framework, guided by the understanding of source code features, which allows the programmer to understand the energy usage of the code and then apply targeted refactoring to save energy. To the best of our knowledge, our work is the first that achieves this for a high-level language such as Java. This approach can be applied at the end of the software engineering implementation phase in order to avoid distracting developers from guaranteeing the correctness of software. 
Energy efficiency can have a significant influence on user experience of mobile devices such as smartphones and tablets. Although energy is consumed by hardware, software optimization plays an important role in saving energy, and thus software developers have to participate in the optimization process. The source code is the interface between the developer and hardware resources. In this paper, we propose an energy-optimization framework guided by a source code energy model that allows developers to be aware of energy usage induced by the code and to apply very targeted source-level refactoring strategies.
The framework also lays a foundation for the code optimization by automatic tools.
To the best of our knowledge, our work is the first that achieves this for a high-level language such as Java.
%however,  effective energy optimization relies mainly on developers, since compiler-based energy optimizations are limited. In this paper, we propose an energy-optimization framework, guided by the understanding of source code features, which allows the programmer to understand the energy usage of the code and then apply targeted refactoring to save energy. To the best of our knowledge, our work is the first that achieves this for a high-level language such as Java. This approach can be applied at the end of the software engineering implementation phase in order to avoid distracting developers from guaranteeing the correctness of software. 
In a case study, the experimental evaluation shows that our approach is able to 
save from 6.4\% to 50.2\% of the CPU energy consumption in various application scenarios.    

	\end{abstract}
	

	% \fi
	%\IEEEpeerreviewmaketitle

	\section{Introduction}

	Smartphones have become widespread in modern society, with a market penetration of about 75\% of mobile subscribers in the U.S in February 2015 \cite{Report:smartphonepenetration}, a figure that is still growing. With the improvement of hardware processing capability and software development environments, the smartphone is no longer just a handset to make phone calls, but
also lets the user play entertaining games, watch movies, browse web pages, and so on. 
However, users are often frustrated by limited battery capacity -- applications running in parallel could easily drain a fully-charged battery within 24 hours -- and therefore energy optimization of applications is of increasing importance. 
	
	Although energy is ultimately consumed by hardware, it is the software that controls the hardware and is often responsible for significant waste of energy. 
Software optimization by current compilers achieves very little energy saving for mobile devices, since besides energy efficiency, the compiler for the mobile device has to consider many other
 important factors, such as limited memory usage and fast response to user interactions. The Android platform, for instance, employs the Just-In-Time (JIT) compiler \cite{justintime}, also known as the dynamic compiler. Its optimization window is generally as small as one or two basic blocks in order to use less memory and speed up delivery of performance boost. 
However, the small window restricts the space of energy-saving strategies. 
Recently, researchers have proposed tools to systematically automate software improvement \cite{GeneticImprovement,Seeds}, but it is hard for developers to guide such optimizations. Powerful code refactoring is needed, but this is beyond the scope of compilers and present tools, relying more on developers' knowledge of the code.  
	 %at levels, such as the algorithm and design,
	%In many cases, code optimization should refer to developers.
	
	Unfortunately, current software development is performed in an energy-oblivious manner and few developers and designers have any awareness of the energy usage of code written by themselves. However,  energy-aware programming techniques are in high demand among software developers. In the most popular software development forum \textsc{StackOverFlow} \cite{stackoverflow}, energy-related questions are marked as favorites 3.89 more often than the average questions \cite{Pinto_miningqusetion}. Furthermore, among energy-related questions, code-design-related ones are prominent. 
	Source code is the interface between the developer and hardware resources; only if developers understand the energy characteristics of the source code can they perform more targeted refactoring to reduce energy use. 
	%Moreover, it has been estimated that energy-saving by a factor of as much as three to five could be achieved solely by software optimization \cite{Edwards:lssmlps}. 
	To realize this goal, the first step is to analyze the source code at different levels of granularity and from different points of view.%energy attributes of
	%The energy information on blocks or more coarse-grained units could identify the hot spots in the code, but it gives few clues about how to make changes to improve the code. The source line is also not an appropriate level of granularity to provide energy information. For instance,  the header of \texttt{for} loop contains three segments which are \textit{initialization}, \textit{boolean} and \textit{update} in the same source line, but usually have distinct numbers of executions. 
	%So the energy information about the source line of the header is not quite appropriate for developers.   

We construct a source-level energy model based on "energy operations", which is fine-grained and gives valuable information for code optimization. %Rather than coarse-grained techniques, this model can distinguish energy consumption of different operations in the same source line.   
By "source-level" we mean that the energy costs of running a program are all attributed to source code constructs, despite the fact that much of the energy consumed is actually accounted for by
things outside the source code such as the operating system.  Thus the model is bound to be an approximation, yet as our results show, it is precise enough to give useful information and guide energy optimization.

	Compared with coarse-grained techniques \cite{Dong_selfconstructivemodel,Zhang_onlinepowerestimation,Wang_batterytrace,sourceline_energy} at the level of source-lines, methods, applications or even the system, there are some advantages of the operation-based model in guiding energy-aware programming techniques:
\begin{figure*}
	\centering
	\includegraphics[width = 0.7\textwidth]{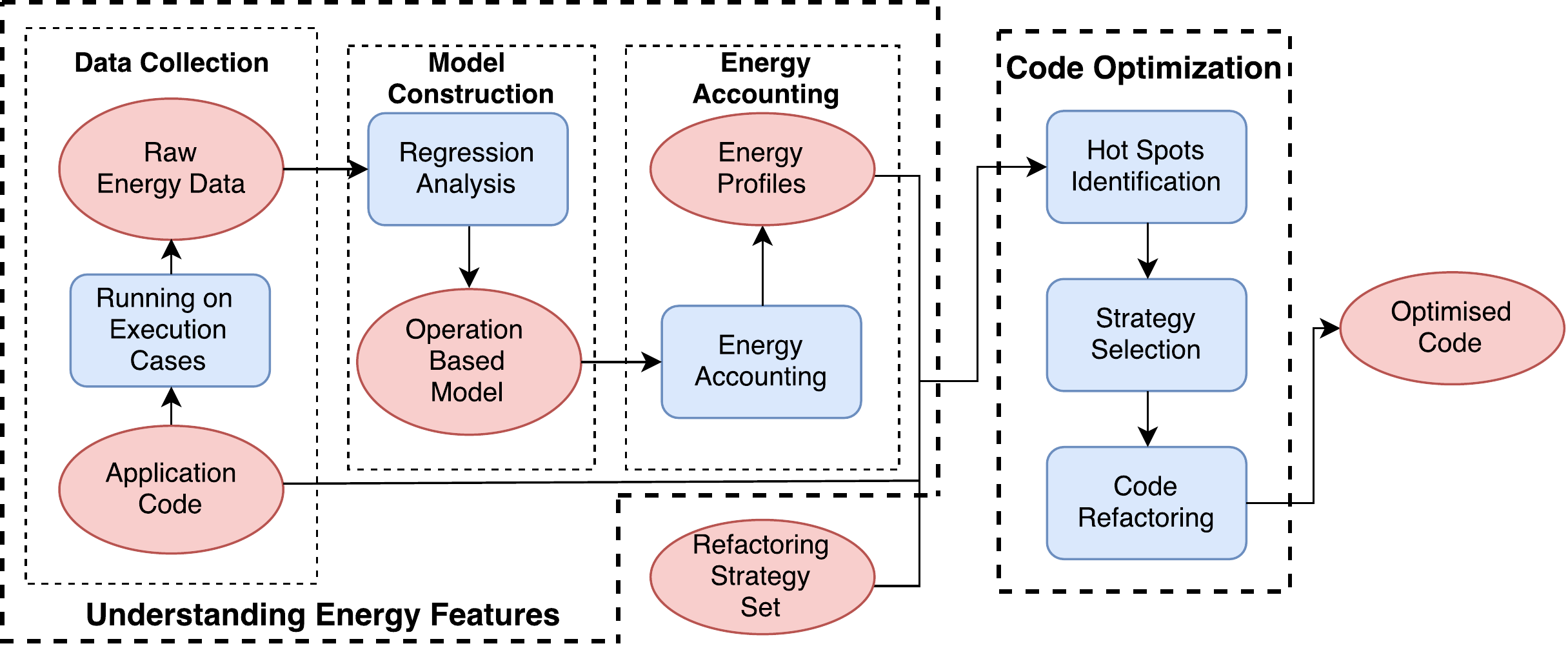}
	\caption{A Framework for Source-Level Energy Optimization }\label{fig:framework}
\end{figure*}	

	\begin{itemize}
		\item The energy operations are atomic units that comprise the entire energy consumption of the application. Thus using the energy estimate of operations,  developers can quantitatively assess the effects of code changes on the energy consumption of code.  	
		%calculate the energy consumption before and after transforming the code, which directs developers on how to make changes.
		
		\item It provides more valuable information for selecting strategies. For example, the experiment shows that method invocation is one of the most expensive operations, suggesting that in some cases we may inline some thin methods, at the cost of losing the integrity of the structure of code. 
	\end{itemize}
	
\iffalse	
	In this paper, we propose an energy-aware programming approach guided by a fine-grained energy model of source code. The summary procedure of the approach is the following:
	
	\begin{itemize}
		\item We build an operation-based source-level energy model, which is achieved by analyzing the data produced in a range of well-designed execution cases.
		\item We perform energy accounting based on the model, at operation and block level to capture the key energy characteristics of the code. 
		\item We focus efforts on the most costly blocks, where we refactor the code to remove, reduce or replace the expensive operations, while maintaining its logical consistency with the original code. 
	\end{itemize} 
\fi

 In this paper, we propose a generic energy optimization framework: 1) understanding the energy features of the source code and 2) optimizing the source code. We then implement one instantiation of the framework, which is guided by an operation-based energy model. 
Briefly, the steps are: 1) we build an operation-based source-level energy model, which is achieved by analyzing the data produced in a range of well-designed execution cases; 2) we perform energy accounting based on the model, at operation and block level to capture the key energy characteristics of the code; 3) we focus efforts on the most costly blocks, where we refactor the code to remove, reduce or replace the expensive operations, while maintaining its logical consistency with the original code.

 The contributions of this paper are the following:
 
  \begin{itemize}
  	\item An energy optimization framework, which is driven by the understanding of energy features of the source code, while the current systematic solutions do not analyze the energy features of the source code before optimizing it. 
  	\item An instantiation of the framework, which builds the infrastructure for software optimization by both automatic tools and developers. 
  	\item An instantiation of the framework guided by an operation-based model. The model can map the energy use to the basic operations at source-level, which is critical to guide optimization. In contrast, traditional profile-based code-optimization techniques \cite{Simunic:2000:source_code_optimization} cannot easily yield operation-related information.
  	%produce operation-level information which is critical in deciding how to refactor the code.   %which can produce more fine-grained energy information than prior profiling techniques.  
  	\item  The evaluation is implemented on a physical device and a real-world game engine. The experimental result shows that the improved code can save the CPU energy consumption by up to 50.2\%.      
  \end{itemize}

	In the rest of this paper, we start with the description of the energy optimization framework in Section \ref{sec:framework}, then introduce the
	identification of source-level energy operations in Section \ref{Section_basicEnergyOp}. In Sections \ref{Section_experimentsetup} and \ref{Section:Model}, we demonstrate the setup and construction of the energy model.  Based on the model we are able to capture energy characteristics and optimize the source code in a case study of three different scenarios, as seen in Sections \ref{Section_clickmove}, \ref{Section_orbit} and \ref{Section_waves} respectively. 
	
%(to make the paper self-contained)

\section{Framework}	\label{sec:framework}

The generic framework of our approach begins with \textit{Understanding Energy Features} of the source code, based on which we perform the \textit{Code Optimization}. Even though it is simple to state, it is a novel approach since the state-of-the-art solutions \cite{GeneticImprovement,Seeds} for systematic code optimization usually treat the code as a black box (i.e., without an analysis of energy features of the code as the first step). Furthermore, realizing the connections between understanding and optimization of code is not trivial.

In this section, we firstly propose a generic framework for source-level energy-optimization, and secondly present an instantiation of the framework to show one way to associate energy features with effective code optimization.       

Figure \ref{fig:framework} shows an overview of the framework. In the rest of this section, we will present and discuss the details.  
	
\subsection{Understanding Energy Features}
There are various ways to analyze and understand the energy characteristics of the source code, and produce a diversity of energy information forming the basis for code optimization. For instance, static analysis for cost upper-bound \cite{CostUpperBound} and computational complexity \cite{ComplexityAnalysis} can give an indication of how good or bad the code is, which may inspire the developers to improve the code. But this high-level information can hardly give clear guidance to developers on how to refactor the code, neither can it facilitate the automation of code optimization.   

Our instantiation of the framework uses an operation-based technique (the explanation of "operation" is in Section \ref{Section_basicEnergyOp}) to analyze the code and provide energy information at a wide range of levels. As shown in Figure \ref{fig:framework}, \textit{Understanding Energy Features} includes three components: \textit{Data Collection}, \textit{Model Construction} and \textit{Energy Accounting}. 

 The \textit{Raw Energy Data} is produced during \textit{Data Collection} by running the application code on a set of well-designed execution cases. \textit{Model Construction} utilizes the \textit{Raw Energy Data} as input to build the operation-based model, powered by which \textit{Energy Accounting} generates energy profiles to capture the key features of the source code. Lastly, the profiles are applied as one crucial input to the \textit{Code Optimization} component.  

\subsection{Code Optimization}

 \textit{Code Optimization} has three main inputs: application code, energy profiles and a set of refactoring strategies. %Basically, the assignment of refactoring strategies is determined by the energy profiles and other properties (such as the code structure, data and control flows, and etc.). 
 Our instantiation is given by the procedure \textit{Code Optimization} described by the pseudo code in Algorithm \ref{OptiAlgorithm}.  

\begin{algorithm}
	\SetKwInOut{Input}{Input}
	\SetKwInOut{Output}{Output}
	
	%\underline{function Euclid} $(a,b)$\;
	\Input{Application Code: $\rho$ \\Energy Profiles (for each code block $id$):\\ \qquad $Blocks=\{(id, cost, \OpCosts) \}$ \\Set of Refactoring Strategies:\\ \qquad $Strategies = \{strategy_i\}$ }
	\Output{Optimized Code: $\rho^\prime$}
	\BlankLine
	\tcp{ Hot Spots Identification (HSI)}
	$HotBlocks = \mathit{findCostlyBlocks}(Blocks)$\;
	\ForEach{$block$ \textbf{in} $HotBlocks$ }
	{
		\tcp{Strategy Selection (SS)}
		$StrategiesToUse = selectStrategies(Strategies,\rho, block)$\;
		\tcp{Code Refactoring (CR)}
		$\rho^\prime = \mathit{applyStrategiesToCode}(\rho, block, \mathit{StrategiesToUse}) $
	}
	
	\caption{The Algorithm of Code Optimization}\label{OptiAlgorithm}
\end{algorithm}

\textbf{Input: } \textit{Application Code} $\rho$ is the source code that developers want to improve.

\textit{Energy Profiles} are the connection between understanding and optimization of source code, and presented in the data structure $Blocks=\{(id, cost, \OpCosts)\}$; $Blocks$ is a set of triples where $id$ is the identity of the block, $cost$ is the energy consumption of the block, $\OpCosts$ is a set containing the energy usage of individual operations in the block.

	\begin{table}
		\small
		\centering
		\caption{Examples of Code Refactoring Strategies\label{RefactorStrategies}}
		\begin{tabular}{ll} 
			\hline
			Category & Examples\\\hline
			& \textit{Loop unrolling, Loop unswitching}\\
			\multirow{-2}{*}{Control Flow} & \textit{If combination, Method inline}\\
			\hline
			& \textit{Common sub-expression elimination} \\
			& \textit{Constant folding and propagation} \\
			& \textit{Loop-invariant code motion}\\
			\multirow{-4}{*}{Data Flow}& \textit{Induction variable elimination} \\
			\hline
			%& \textit{Loop-invariant code motion}\\
			%& \textit{Method inline}\\
			Other & \textit{Replacement by library function} \\
			\hline
		\end{tabular}
	\end{table}

The \textit{Set of Refactoring Strategies} represents the available optimization strategies to apply to the code. Table \ref{RefactorStrategies} lists examples of  code refactoring strategies applied in our case study. 
%The objective of strategies is to remove, reduce or replace the expensive operations. 
Usually, one strategy is targeted to reduce one category of operations, which facilitates automatic selection of strategies. For example, if arithmetic operations are the major energy consumers, the data-flow strategies are likely to be chosen; if method invocations are costly, then method inlining is selected. Additionally, replacement of a source-level function by a library function reduces energy consumption for both data and control related operations since the latter is already compiled into native code which does not incur costly run-time compilation.  
%

%The examples of the strategies are shown in Table \ref{RefactorStrategies}.The case study in Section \ref{Section_clickmove}, \ref{Section_orbit} and \ref{Section_waves} adopts the strategies, like method inline, loop unrolling and replacement to library function, which are shown effective to reduce the costly operations such as method invocations and jumps from block to block (these operations are expensive in the application source code, which is proved in the experiment).  
%

The \textit{Code Optimization} algorithm consists of \textit{Hot Spots Identification (HSI)}, \textit{Strategy Selection (SS)} and \textit{Code Refactoring (CR)}.

\textit{HSI} is a component that shows clearly which parts are the most energy-consuming and suggests to the optimization tools or developers where to focus efforts, rather than examining the whole application code. 
As seen in Algorithm \ref{OptiAlgorithm}, \textit{HSI} is implemented by $findCostlyBlocks()$ which identifies the costly blocks ($HotBlocks$) by comparing the $cost$ element in their triples.  
There are several possible implementations of $findCostlyBlocks()$. For instance it could compute all the blocks in order of energy cost, choosing the most costly as the $HotBlocks$. Another approach is to return blocks that consume more than say 10\% of the overall energy usage as the $HotBlocks$.

Thereafter, the algorithm traverses the $HotBlocks$ and for each block, the \textit{SS} component chooses the refactoring strategies according to several criteria: the energy breakdown on operations (referring to \textit{OpCosts} in each \textit{block}), the structural characteristics of the code around and within that \textit{block}. For example, if control-flow operations are the most costly and the \texttt{for} loop is the syntax structure around the block, \textit{SS} will adopt loop unrolling for refactoring. 

Finally, \textit{CR} applies the selected strategies to improve the code. The scope in which the code is refactored is not limited within a block; i.e. some code outside the block may be changed to reduce the costly operations in the block. The example strategy for this case is loop unrolling.  $\rho^\prime$ is the  \textit{Optimized Code}  resulting from the refactoring. 
In summary, this algorithm presents a systematic approach to optimizing the source code, providing a framework for optimization by both automatic tools and developers.

 For simple strategies, like loop unrolling and method inlining, refactoring can be automated. Even though some of these simple strategies could be done by other tools such as the compiler, focusing on $\mathit{HotBlocks}$ can be crucial. For instance, method inlining or loop unrolling are important energy-saving techniques, but if applied indiscriminately, the size of code can explode.

For algorithm and design level strategies, such as replacement of a source-level function by a library function, the refactoring relies more on the developers because it is very difficult for automatic tools to identify which pieces of code can be replaced by library functions. In fact, \textit{Energy profiles} are completely human-readable, and \textit{HSI} reduces the effort needed for manual code improvement, which is shown in the case study in Section \ref{Section_clickmove}, \ref{Section_orbit} and \ref{Section_waves}.

	\section{Basic Energy Operations}\label{Section_basicEnergyOp}

	The operational semantics of a language specifies the effect of each language construct on the behavior of a program.  Guided by the semantics of the source code, we select a set of \emph{energy operations}, which are basic constructs such as statements and functions that cause evaluation or state changes. We assume that the energy consumed during program execution can be attributed to these, and only these operations.  The choice of energy operations is thus an informed guess;  operations not appearing in the source code that consume energy will not be directly reflected in the energy model (examples could be garbage collection or operating system tasks);  their energy will be absorbed into the source code operations. On the other hand, if we select operations that have little or no energy effect, this will not cause problems as they will automatically be identified by the regression analysis in the later stage of the analysis.
	
	Our experiment focuses on the Java language for which an operational semantics is available \cite{Bogdanas_Semantics} to inspire the selection of Java source-code energy operations. Table \ref{EnOps} lists 14 representative operations out of a total of 120 in the experiment, giving them names that correspond to their function and argument types.
	They include arithmetic calculations like \textit{Multi\_float\_float}, \textit{Addition\_int\_int}, in which operands types are explicit, as well as \textit{Increment} whose operand is implicitly an integer.   Boolean operations and comparisons, such as \textit{And}, \textit{Less\_int\_float} and \textit{Equal\_Object\_null} also form a major category. \textit{Method Invocation} and \textit{Block Goto} are important for the control flow which plays a key role in the execution of the code. Assignments and \textit{Array Reference} are expensive, as will be shown in Section \ref{Section_Analysis}. %take a significant amount of the application's energy consumption, as will be shown in Section \ref{Section_Analysis}. 
	
\begin{table}
		\small
		\centering
		\caption{Examples of Energy Operations\label{EnOps}}
		\begin{tabular}{ll} 
			\hline
			Operation &  \textit{Identified where:} \\ \hline%\hline
			%\multicolumn{1}{c}{}
			%\rowcolor{Gray}
			{Method Invocation} & \textit{\small one method is called}\\ %\hline
			
			Parameter\_Object &  \textit{\small Object is one parameter of the method}\\ %\hline
			%&  Identified when``Object" is a method parameter \\	
			%\multirow{-2}{*}{Parameter\_Object} &  of the method \\
			
			Return\_Object &  \textit{\small the method returns an Object}\\ %\hline
			
			%\cellcolor{Gray} & \cellcolor{Gray}Identified when symbol ``$+$" appears in code,  \\  
			%\multirow{-2}*{ \cellcolor{Gray} Addition\_int\_int}& \cellcolor{Gray}and two operands are integers\\ 
			%\rowcolor{Gray} 
			Addition\_int\_int & \textit{\small addition's operands are integers }\\ %\hline

			Multi\_float\_float & \textit{\small multiplication's operands are floats }\\
			%\hline
			%\rowcolor{Gray}
			Increment & \textit{\small symbol "$++$" appears in code}\\ 
			%\hline
			And & \textit{\small symbol "$\&\&$" appears in code}\\ 
			%\hline
			%\rowcolor{Gray}
			Less\_int\_float & \textit{\small "<"'s operands are integer and float}\\%\hline
			%\cellcolor{Gray} & \cellcolor{Gray}Identified when symbol ``$<$" appears in code,  \\  
			%\multirow{-2}*{ \cellcolor{Gray} Less\_int\_float}& \cellcolor{Gray}and one operand is integer, %one is floating\\  
			
			Equal\_Object\_null & \textit{\small "=="'s operands are Object and null}\\ %\hline
			%&  Identified when symbol ``$==$" appears in code, \\	
			%\multirow{-2}{*}{Equal\_Object\_null} &  and one operand is ``Object", one is ``null" \\
			
			%\rowcolor{Gray}
			Declaration\_int & \textit{\small one integer is declared}\\%\hline
			
			Assign\_Object\_null & \textit{\small assign  operands are Object and null}\\
			%\hline
			%	&  Identified when symbol ``$=$" appears in code, \\	
			%	\multirow{-2}{*}{Assign\_Object\_null} &  and `null" is assigned to an ``Object" \\	
			%\rowcolor{Gray}	 
			Assign\_char[]\_char[] & \textit{\small assign operands are arrays of chars}\\%\hline
			%\cellcolor{Gray} & \cellcolor{Gray}Identified when symbol ``$=$" appears in code,  \\  
			%\multirow{-2}*{ \cellcolor{Gray} Assign\_char[]\_char[]}& \cellcolor{Gray}and operands are two %arrays of chars \\  
			
			Array Reference &  \textit{\small one array element is referred}\\
			%\hline
			%	\rowcolor{Gray}	 
			Block Goto & \textit{\small the code execution goes to a new block}\\
			%\cellcolor{Gray} & \cellcolor{Gray}Identified when the code execution goes to  \\  
			%\multirow{-2}*{ \cellcolor{Gray} Block Goto}& \cellcolor{Gray}a new block \\ 

			\hline
		\end{tabular}
	\end{table}

\iffalse
	\begin{table}
		\small
		\centering
		\caption{Examples of Library Functions\label{Libaray_functions}}
		\begin{tabular}{ll} \hline
			Class & $\qquad$Function \\ 
			\hline%\hline
			%\rowcolor{Gray}
			ArrayList & \textit{add, get, size, isEmpty, remove}   \\ %\hline
			
			& \textit{glBindTexture, glDisableClientState } \\
			& \textit{glDrawElements, glEnableClientState} \\
			GL10     & \textit{glMultMatrixf, glTexCoordPointer} \\
			& \textit{glPopMatrix, glPushMatrix} \\
			& \textit{glTexParameterx, glVertexPointer}\\
			%\hline
			%\rowcolor{Gray}
			Math    & max, pow, sqrt, random  \\%\hline
			
			FloatBuffer  & \textit{position, put} \\
			\hline
		\end{tabular}
	\end{table}
\fi	
	Applications often employ a diversity of library functions, some of which are frequently called (graphics functions, for example). Unlike normal code which is interpreted by a virtual machine at run-time,  key parts of library code has been compiled into native code before execution and some part may be already written in different languages and at lower levels of the software stack. Thus we include library functions as energy operations.% E.g., the functions in class \textit{GL10} are responsible for graphic computing. 

	\section{Experimental Setup}\label{Section_experimentsetup}
	
	In this section and the next, we summarize the construction of the energy model for a generic class of Android applications based on a game engine, including the setup of the target device and the design principles of the execution cases. Further details on these can be found in \cite{Xueliang_modeling}. Note that, this setup is also applied to the evaluations of the code refactoring, as seen in Section \ref{Section_candmEvaluation}, \ref{orbit_evaluation} and \ref{waves_evaluation}. 
	
	\subsection{Experimental Targets}\label{Section_target_measurement}
	\textbf{Device.} We employ an Odroid-XU+E development board \cite{target:odroid} as the target device. It possesses two ARM quad-core CPUs, which are Cortex-A15 with 2.0 GHz clock rate and Cortex-A7 with 1.5 GHz. 
	Odroid-XU+E has built-in sensors to measure the voltage and current of CPUs. These sensors are supposed to be integrated into the future architecture of mobile devices since they provide the ground truth for run-time energy modeling and optimization.   

	%The eight cores are logically grouped into four pairs. Each pair consists of one big and one small core. So from the operating system's point of view there are four logic cores. 
	In our experiment, we turn off the small cores (because in several execution cases, small cores cannot afford the workload) and run workload on big cores at a fixed clock frequency of 1.1 GHz. We do this also in order to control the influence of voltage, clock rate and CPU performance on energy usage because we are only concerned about the effects of basic operations.   
	 	
	 %with a frequency of 30 Hz. %and updates the samples in a log file. 
	
	%We wrote a script to obtain the samples from the file. During execution we run the script on an idle core to minimize its influence on the application.   
	
	%Note that the power monitor gives two sequences of power samples: one is for the big cores and the other is for the small cores. We pick the sequence of power samples of the big cores, because we only run workload on them.
	
	%the power reading has two segments: one is the power sample of four big cores, the other is the power sample of four small cores. We pick the big cores' power sample, because we only run workload on big cores.

	\textbf{Application Source Code.}
	The target source code is the Cocos2d-Android \cite{code:cocos2d} game engine, a framework for building games, demos and other interactive applications such as virtual reality. It also implements a fully-featured physics engine. Games are increasingly popular on mobile phones and include more and more fancy and energy-consuming features, requiring high CPU performance. Our instantiation of the framework demonstrates the energy modeling, accounting and improvement for the source code of the game engine, and evaluates the improvement in three game scenarios. 
	
	%but the methodology of energy modeling, accounting and code optimization is applicable for all kinds of applications.
	
	\subsection{Design of Execution Cases }\label{Section_sourcecode_casedesign}
	
	The execution cases whose energy usage is measured and analyzed represent typical sequences of actions during  game, including user inputs. We focus on three scenarios which are \texttt{Click \& Move}, \texttt{Orbit} and \texttt{Waves}.

	In the \texttt{Click \& Move} scenario, the sprite (the character in the game) moves to the position where the tap occurs. In the \texttt{Orbit} scenario, the sprite together with the grid background spins in the three-dimension space. In the \texttt{Waves} scenario, the sprite scales up and down, meanwhile the grid background waves like flow. In both the \texttt{Orbit} and \texttt{Waves} scenarios, the animation will restart from the starting point whenever and wherever the tap occurs. 
	
	To simulate the game scenarios under different sequences of user inputs, we script with the Android Debug Bridge \cite{adb:android} (ADB), a command line tool connecting the target device to the host, to automatically feed the input sequences to the target device.
	
	An execution case is made up of one user input sequence and one set of basic blocks. In order to obtain a more varied set of execution cases and thus a more precise model, we vary the executions of individual basic blocks in the code. This is achieved by systematically removing a set of blocks for each execution case, using the control flow graph extracted using the Soot tool \cite{soot:callgraph}. We ensure that each block could be removed in some execution case and thus execution sequences are not restricted to ``normal" behavior but contain some randomness.
		%The problem is that a certain amount of blocks are critical to the functionality of the game engine, so we avoid removing them in the design stage.    

\subsection{Energy Consumption from Power samples}
     Power is equal to voltage times current (voltage and current are obtained from the sensors); we approximate energy consumption by calculating Equation (\ref{Energy_Equation}): $p=\emph{power}(t)$ is the power trace, that is, the continuous power-vs-time function; $\emph{power}(t_i)$ is the power sample at time-stamp $t_i$; $\Delta_i$ equals to $t_i - t_{i-1}$, which is the interval between two consecutive samples.

    \begin{equation}\label{Energy_Equation}E=\int_{t_0}^{t_n}\textit{power}(t)\,dt
    \;\approx\sum_{i=1}^{n}\textit{power}(t_i)\cdot\Delta_i\; \end{equation}
    
    \begin{displaymath}\quad where
    \quad t_0\le t_1\le t_2 \cdots \le t_{n-1}\le t_{n}  \end{displaymath}
    
    \textbf{Control of Measurement Variability.} We run each execution case 10 times (when the cooling fan keeps the CPU temperature stable at 51$^{\circ}{\rm C}$) to obtain 10 records of the energy consumption computed by Equation \ref{Energy_Equation}. We use the coefficient of variation ($C_v$) to represent the variability of the records. $C_v$ is computed by $C_v = \frac{\sigma}{\mu}$, where $\sigma$, $\mu$ are the standard deviation and mean of the 10 records for each execution case. The experimental results show that the mean of $C_v$s of all the execution cases is about 1.6\%, indicating that the variability is very limited. We employ the mean of 10 records as the "real" energy consumption of the execution case, which makes the variability as small as 0.5\%.% Note that, the values of energy usage shown in the evaluation stage (Section \ref{Section_candmEvaluation} \ref{orbit_evaluation} and \ref{waves_evaluation}) are also acquired in this way.     

\section{Model Construction}\label{Section:Model}

%The aimed model is formalized in Equation \ref{Energy_Model}. The entire energy consumption consists of the sum of the costs of operations and library functions and idle cost. Notice that the idle costs of individual cases are different, since they are executed in distinct sequences of inputs, and the lengths of sessions are also varying. So we measure the idle cost for individual cases.   
The entire energy use is composed of three parts: the cost of energy operations, the cost of library functions and the idle cost. The model is formalized in Equation (\ref{Energy_Model}):

\begin{equation}\label{Energy_Model}
E = \sum_{}^{op_i \in Energy\,Ops} Cost_{op_i} \cdot  N_e(op_i) \qquad\qquad \end{equation}
\begin{displaymath}
+ \sum_{}^{func_i \in Lib\,Funcs} Cost_{func_i} \cdot  N_e(func_i) + Idle\;Cost
\end{displaymath}

 The cost of energy operations is the sum of $  Cost_{op_i} \cdot  N_e(op_i) $ (the cost of one operation multiplied by the number of its executions), where $op_i \in Energy\,Ops$,  the set containing all the operations. The cost of library functions is the sum of $Cost_{func_i} \cdot  N_e(func_i)$ (the cost of one library function multiplied by the number of its executions), where $func_i \in Lib\,Funcs$; $Lib\,Funcs$ is the set of library functions.
The $Idle\; Cost$ is the energy consumption of the device when running no application, but simply the Android system.% The lengths of case sessions are diversified due to input sequences, so the $Idle\; Cost$ is different for each execution case.
%The model construction is based on regression analysis, finding out the correlation between energy operations and their costs from the data obtained in the execution cases. 

Model construction is based on regression analysis. Each execution case produces one example for training the model, whose purpose is to capture the correlation between energy operations and their costs from the examples produced by all the execution cases. 
%The $N_e(op_i)$, $N_e(func_i)$, $Idle\; Cost$ and $E$ are obtained from measurement or tracing. $Cost_{op_i}$ and $Cost_{func_i}$ are the values to estimate.
\iffalse
We set out the collected data in the matrices in Equation (\ref{equation_matrices}). The leftmost matrix ($N$) contains the execution numbers of $l$ operations (including energy operations and library functions) in $m$ execution cases. Each row indicates one execution case. Each column represents one operation. The vector ($\vec{cost}$) in the middle contains the costs of $l$ operations, which are the values we are aiming to estimate. The vector ($\vec{e}$) on the right of the equal mark contains the measured entire energy costs of the execution cases. So for each execution case, the entire energy cost is the sum of the costs of operations. It should be noticed that the energy costs $\vec{e}$ exclude the $Idle\; Cost$ which is measured when no application workload is being processed. Note that, $N$ and $\vec{e}$ are collected data, $\vec{cost}$ is the one to estimate.

\begin{equation}\label{equation_matrices}
\begin{pmatrix} 
n_1^{(1)}  & n_2^{(1)} & ... & n_l^{(1)}\\ 
n_1^{(2)}  & n_2^{(2)}  & ... & n_l^{(2)}\\
&        ...& ...    &          \\
n_1^{(m-1)}  & n_2^{(m-1)}  &  ... & n_l^{(m-1)}\\
n_1^{(m)}  & n_2^{(m)}  &  ... & n_l^{(m)}
\end{pmatrix} \times
\begin{pmatrix}
cost_1\\
cost_2\\
...\\
cost_l
\end{pmatrix} =
\begin{pmatrix}
e_1\\
e_2\\
...\\
e_{m-1}\\
e_m
\end{pmatrix}\\
\end{equation}
\fi
To validate the model, we apply a four-fold cross validation procedure: the set of execution cases is randomly evenly divided into four subsets; in each one of  four rounds in all, one of the subsets is chosen to be the validation set and the others together to be the training set. We utilize two statistical criteria to assess our model. The first one is the correlation coefficient ($r$) that represents the strength and direction of the linear relationship between estimated and measured values. The result shows $r$ in the training sets is from 0.81 to 0.84, and the validation sets from 0.88 to 0.91 which means the estimated value has a positive and strong relationship with its corresponding measured value. 

The other criterion is the Normalized Mean Absolute Error (NMAE). The NMAE is a well-known statistical criterion that indicates how well the estimated value matches the measured one. It is computed by Equation (\ref{equation_NMAE}), the mean value of normalized difference between the predicted energy cost $\hat{e}$ and the measured cost $e$. The lower the ratio the better the result. The NMAE in training sets ranges from 14.1\% to 16.3\%, and in validation sets from 9.3\% to 15.7\%. 
%The NMAEs are around 15.0\%, which means the model's inference accuracy is around 85.0\%.

\begin{equation}\label{equation_NMAE}
NMAE= \frac{1}{n} \sum_{i=1}^{n} | \frac{\hat{e^{(i)}} - e^{(i)}}{e^{(i)}} |
\end{equation}

We choose the model that performs well (accuracy above 85\%) both in training and validation sets for the later energy accounting and code optimization stage.

Next we present three instances of the code optimization procedure based on the energy model, applying it to code for performing three typical game scenarios:  \texttt{Click \& Move},  \texttt{Orbit} and \texttt{Waves}. The description of the scenarios is shown in Section \ref{Section_sourcecode_casedesign}.

\section{The \texttt{Click \& Move} Scenario}\label{Section_clickmove}

In this section, we first discuss the operation costs computed by the energy model, after which we apply the energy optimization procedure from Algorithm \ref{OptiAlgorithm}, consisting of hot spot identification, strategy selection and code refactoring. Lastly, we present an evaluation  of the refactoring strategies.   

%begin with energy accounting at operation and block level for the  \texttt{Click \& Move} scenario, after which we improve the most costly blocks focusing on the most expensive operations. We apply a similar approach in the other scenarios \texttt{Orbit} and \texttt{Waves} in Section \ref{Section_orbit} and Section \ref{Section_waves}; however for those cases we will only briefly summarize energy accounting and focus on the code improvements. 

\subsection{Operation Costs}\label{Section_Analysis}
%The energy model of app source code based on energy operations facilitates comprehensive energy accounting from operation-level up to source-level. In this section, we will see the rank of the most expensive operations, and the contributions of different operations to the energy consumption of each block.  

\begin{figure}
	\centering
	\includegraphics[width = 0.44\textwidth]{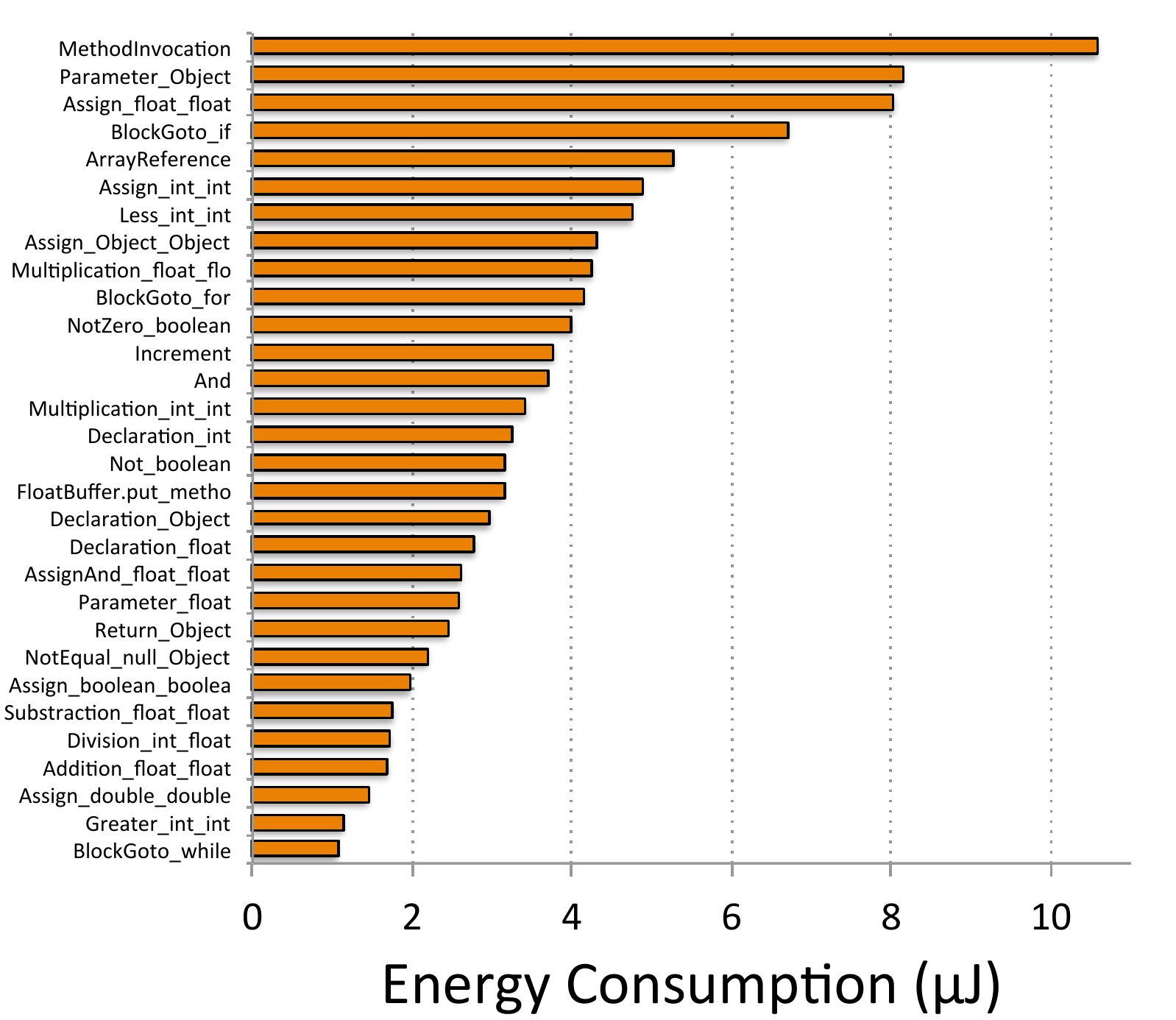}
	\caption{The top 30 energy consuming operations in \texttt{Click \& Move} scenario.}\label{fig:op_rank}
\end{figure}

%\paragraph{Operation Level}
Figure \ref{fig:op_rank} shows the top 30 energy consuming operations in the model, ranked by their single-execution energy costs. 
%The line marked "71.3\% Energy Consumption" indicates the percentage of the energy cost of the execution cases for the \texttt{Click \& Move} scenario contributed by the top 10 operations. Similarly, "26.1\% Energy Consumption" shows the contribution to the total cost of operations from 11th to 30th. We can see that the energy usage of the code is largely determined (97.4\%) by a relatively small  number of operations. This is due to the fact that these operations are frequently used and expensive in themselves.
%The 30 operations out of 187 (including library functions) take up  of the total cost, in which the top 10 consumes the major part with a percentage of 71.3\%.
%Among all the operations (even beyond the top 30), the single-execution costs of them vary in a range of several orders of magnitude. 
It might be supposed that the sophisticated arithmetic operations, such as multiplications and divisions, should be the most costly. However, the result shows that \textit{Method Invocation} ranks the highest. This is due to a sequence of complex processes to fulfill \textit{Method Invocation}, for example, most of the method calls in Java are virtual invocations which are dispatched on the type of the object at run-time and implicitly passed a "this" reference as their first parameter, not to mention other operations such as storing the return address and managing the stack frame.
%and \textit{Parameter\_Object}, two method-relevant operations,

This suggests a trade-off between code structure and energy saving when writing the code. That means, in certain cases, we could inline some thin and highly-invoked methods in the code, at the cost of losing the integrity of the structure of the code to some extent. 

%Only one arithmetic operation, namely \textit{Multi\_float\_float}, is a member of the top 10, and there are only six arithmetic operations in the top 30. They together cost only 6.1\% of the overall energy consumption of the application, which is somewhat unexpected. 

%\textit{Increment}, \textit{Multi\_int\_int}, \textit{Subtraction\_float\_float}, \textit{Division\_int\_float} and \textit{Addition\_float\_float}. 
%Two arithmetic operations, \textit{Addition\_int\_int} and \textit{Multi\_float\_\\float}, are members of the top 10. Unexpectedly, the addition is twice as expensive as the multiplication. We surmise that this is a result of their operands in the target code, as experimentally shown in \cite{Steve_data_dependent}, the energy cost of the arithmetic computation is operand-dependent.

%Later in block-level energy accounting, we will see that assignments, comparisons and \textit{Array Reference} play significant roles in the overall energy consumption. This is not only because they are frequently used, but also because they are costly as operations themselves, as shown in Figure \ref{fig:op_rank}. 

\textit{Block Goto} operations are expensive as well.
Based on the types of conditionals and loops where "Block Goto" occurs, they are classified into \textit{BlockGoto\_if}, \textit{BlockGoto\_for} and \textit{BlockGoto\_while}. The result shows that they cost different amounts of energy as operations themselves, respectively 6.7 $\mu$J, 4.1 $\mu$J and 1.1 $\mu$J. Together with \textit{Method Invocation}, they take up 37.6\% of the total application energy consumption. 

 To facilitate the discussion on operations in the reminder of this paper, we classify a number of operations into groups. Specifically, the "Block Goto" operations, \textit{ Method Invocation} and field references are gathered in \textit{Control Ops}; the parameter passing and the value returns of methods are in \textit{Function Ops}; the comparisons and Booleans are in \textit{Boolean Ops}; all the arithmetic computations are in \textit{Arithmetic Ops}; all the library functions are in \textit{Lib Functions}.

\subsection{Code Optimization}

\iffalse
\begin{table}
	\small
	\centering
	\caption{The top 10 most costly blocks in \texttt{Click \& Move}. \label{table_topblocks}}
	\begin{tabular}{lrr} \hline
		\quad Block ID & \#Executions & Energy Cost (mJ)\\ 
		\hline%\hline
		CCNode.visit() & 19462 \quad\quad & 2128.6 \qquad\quad \\
		CCNode.transform() & 18903 \quad\quad& 1648.4 \qquad\quad\\
		CCTextureAtlas.putVertex() & 2119 \quad\quad & 1494.4 \qquad\quad\\
		CCNode.visit().if\_4.for\_1 & 16880 \quad\quad& 1426.8 \qquad\quad\\
		CCNode.transform().if\_1   & 19664 \quad\quad & 1426.3 \qquad\quad\\
		CCTextureAtlas.putTexCoords() & 2120 \quad\quad&  1107.8 \qquad\quad\\
		CCAtlas.updateValues().for\_1 & 2173 \quad\quad& 1018.7 \qquad\quad\\
		CCNode.visit().if\_3.for\_1 & 8356 \quad\quad & 915.7 \qquad\quad\\
		CCSprite.draw() & 8594 \quad\quad& 766.9 \qquad\quad\\
		CCTexture2D.name() & 13085 \quad\quad& 537.5 \qquad\quad\\
		\hline
	\end{tabular}
\end{table}
\fi
%The most important consideration of app developers is to guarantee the correctness of software, which should then be followed by energy efficiency. So our energy-aware programming approach is adopted at the end of software engineering life circle when the software system is in general complete. 

%The overview of energy-aware programming approach is firstly finding the most costly blocks, where we analyze the energy breakdown on the operations, and make changes to the code to diminish the usage of costly operations.

\subsubsection{Hot Spot Identification}

\begin{table}
	\small
	\centering
	\caption{The top 10 most costly blocks in \texttt{Click \& Move}. \label{table_topblocks}}
	\begin{tabular}{lr} \hline
		\quad Block ID  & Energy Cost (mJ)\\ 
		\hline%\hline
		CCNode.visit()  & 2128.6 \qquad\quad \\
		CCNode.transform() & 1648.4 \qquad\quad\\
		CCTextureAtlas.putVertex()  & 1494.4 \qquad\quad\\
		CCNode.visit().if\_4.for\_1 & 1426.8 \qquad\quad\\
		CCNode.transform().if\_1    & 1426.3 \qquad\quad\\
		CCTextureAtlas.putTexCoords() &  1107.8 \qquad\quad\\
		CCAtlas.updateValues().for\_1 & 1018.7 \qquad\quad\\
		CCNode.visit().if\_3.for\_1  & 915.7 \qquad\quad\\
		CCSprite.draw() & 766.9 \qquad\quad\\
		CCTexture2D.name() & 537.5 \qquad\quad\\
		\hline
	\end{tabular}
\end{table}
In practice, a hot spot is the size of a block. Using the energy profiles ($Blocks=\{(id, cost, OpCosts)\}$),   
we identify the 10 most costly blocks when \texttt{Click \& Move} runs without removing any block (see Table \ref{table_topblocks}). For example, \textit{CCNode.visit()} is the entrance block of the \textit{visit()} function; \textit{CCNode.visit().if\_4.for\_1} is the body block of the \texttt{for} loop.
These 10 blocks are distributed in seven methods, so the code review is straightforward.  

\subsubsection{Strategy Selection \& Code Refactoring}

We find four easy opportunities to improve energy efficiency of some blocks:  \textit{CCNode.visit()}, \textit{CCNode.visit().if\_4.for\_1} and \textit{CCTexture2D.name()}. There are also other opportunities in other blocks supposed possible to save energy, but requiring more efforts and gaining little. For example, \textit{CCAtlas.updateValues().for\_1} has several busy arithmetic expressions. Usually it is assumed that replacing the busy expression with a variable would reduce energy, however in this case the overhead of variable declaration counteracts the saved energy.
%The four opportunities to reform the code are very simple and effective, but can only be discovered by the operation-level information. The changes are shown as follows:

\begin{program}
	\begin{lstlisting}
	if (children_ != null) {
	    if_body1;
	}
	draw(gl);
	if (children_ != null) {
	    if_body2;
	}
	
	\end{lstlisting}
	\caption{ Simplified parts of \textbf{original} code in \textit{CCNode.visit()}}
	\label{program1}
\end{program}

\begin{program}
	\begin{lstlisting}
	if (children_ != null) {
	    if_body1;
	    draw(gl);
	    if_body2;
	} else {draw(gl);}
	\end{lstlisting}
	\caption{The changed Program \ref{program1} }
	\label{program2}
\end{program}

\textit{(i) If Combination: }
This change is made in the most costly block \textit{CCNode.visit()}, which has two comparisons, two Boolean operations, one \textit{Method Invocation} and one parameter passing. In fact, the two \texttt{if} headers make the same comparison, as shown in Program \ref{program1}. We change the code to Program \ref{program2}, which combines the two \texttt{if} statements and meanwhile keep it logically consistent with Program \ref{program1}. By these means each execution of the block can remove one comparison, and when the condition is false, it can additionally remove one \textit{BlockGoto\_if}. 

\iffalse
\begin{program}
	\begin{lstlisting}
	public void visit(GL10 gl) {
	       ......
	    transform(gl);
	       ......
	}
	public void transform(GL10 gl) {
	    tranform_body;
	}
	\end{lstlisting}
	\caption{Simplified parts of \textbf{original} code in \textit{CCNode} class}
	\label{program3}
\end{program}

\begin{program}
	\begin{lstlisting}
	public void visit(GL10 gl) {
	       ......
	    transform_body;
	       ......
	}
	public void transform(GL10 gl) {
	    transform_body;
	}
	\end{lstlisting}
	\caption{The changed Program \ref{program3}}
	\label{program4}
\end{program}

\fi

\textit{(ii) Inner-Class Method Inline: }
When \texttt{Click \& Move} runs with full set of blocks, the \textit{transform()} function is invoked 18903 times and mostly by the \textit{visit()} function.
We switch the body of \textit{transform()} to the function call of \textit{transform()} in \textit{visit()}, meanwhile retaining the original definition of \textit{transform()} in case that other parts of the code call it. This change can greatly decrease the number of calls to \textit{transform()}s and thus \textit{Method Invocation}s that are costly. However, it may be at the cost of losing readability of the code (which might be partly compensated by adding explanatory comments). 

\textit{(iii) Loop-Invariant Code Motion:}
% as seen in Program \ref{program5} , as shown in Program \ref{program6}
\textit{CCNode.visit().if\_3.for\_1} and \textit{CCNode.visit().if\_4.for\_1} are entrance blocks of the two \texttt{for} loops. These two loops share a quantity, \textit{children\_.size()}, which is computed in each iteration but actually constant. We thus hoist it outside the loop, which saves the energy of invoking and executing the \textit{size()} function during every iteration. Meantime, we move the declaration of the \textit{child} outside the loop, considering the cost of \textit{Declaration\_Object} is about 2.97 $\mu$J and also among the top 30 most costly operations. 

\textit{(iv) Inter-Class Method Inline: }
\textit{CCTexture2D.name()} is the 10th most costly block and costs 537.5 mJ when \texttt{Click \& Move} runs with full set of blocks. However, its job is to simply get the value of the private member variable, \textit{\_name}, of the class \textit{CCTexture2D}. This method has only two callers in the code. So we consider to make this variable public and let the two callers directly get access to the variable, which avoids the cost of \textit{Method Invocation}. This change may harm the encapsulation of data, however, only one member of one class is changed. The trade-off between energy saving and data encapsulation will in the end be decided by developers. 

\iffalse
\begin{program}
	\begin{lstlisting}
	if (children_ != null) {
	for (int i=0; i<children_.size(); ++i) { 
	    CCNode child = children_.get(i);
	    if (child.zOrder_ < 0) {
	         child.visit(gl);
	    } else
	        break;
	}
	draw(gl);
	for (int i=0; i<children_.size(); ++i) { 
	    CCNode child = children_.get(i);
	    if (child.zOrder_ >= 0) {
	        child.visit(gl);
	    }
	}
	} else {draw(gl);}
	\end{lstlisting}
	\caption{The full version of Program 2}
	\label{program5}
\end{program}

\begin{program}
	\begin{lstlisting}
	CCNode child = new CCNode(); //added
	int children_size = children_.size(); //added
	if (children_ != null) {
	    for (int i=0; i<children_size; ++i) { //changed 
	        child = children_.get(i); //changed
	        if (child.zOrder_ < 0) {
	            child.visit(gl);
	        } else
	              break;
	    }
	draw(gl);
	for (int i=0; i<children_size; ++i) { //changed
	    child = children_.get(i); //changed
	    if (child.zOrder_ >= 0) {
	        child.visit(gl);
	    }
	}
	} else {draw(gl);}
	\end{lstlisting}
	\caption{The changed Program \ref{program5}}
	\label{program6}
\end{program}

\fi

\subsection{Evaluation}\label{Section_candmEvaluation}

\begin{figure}
	\centering
	\includegraphics[width = 0.4\textwidth]{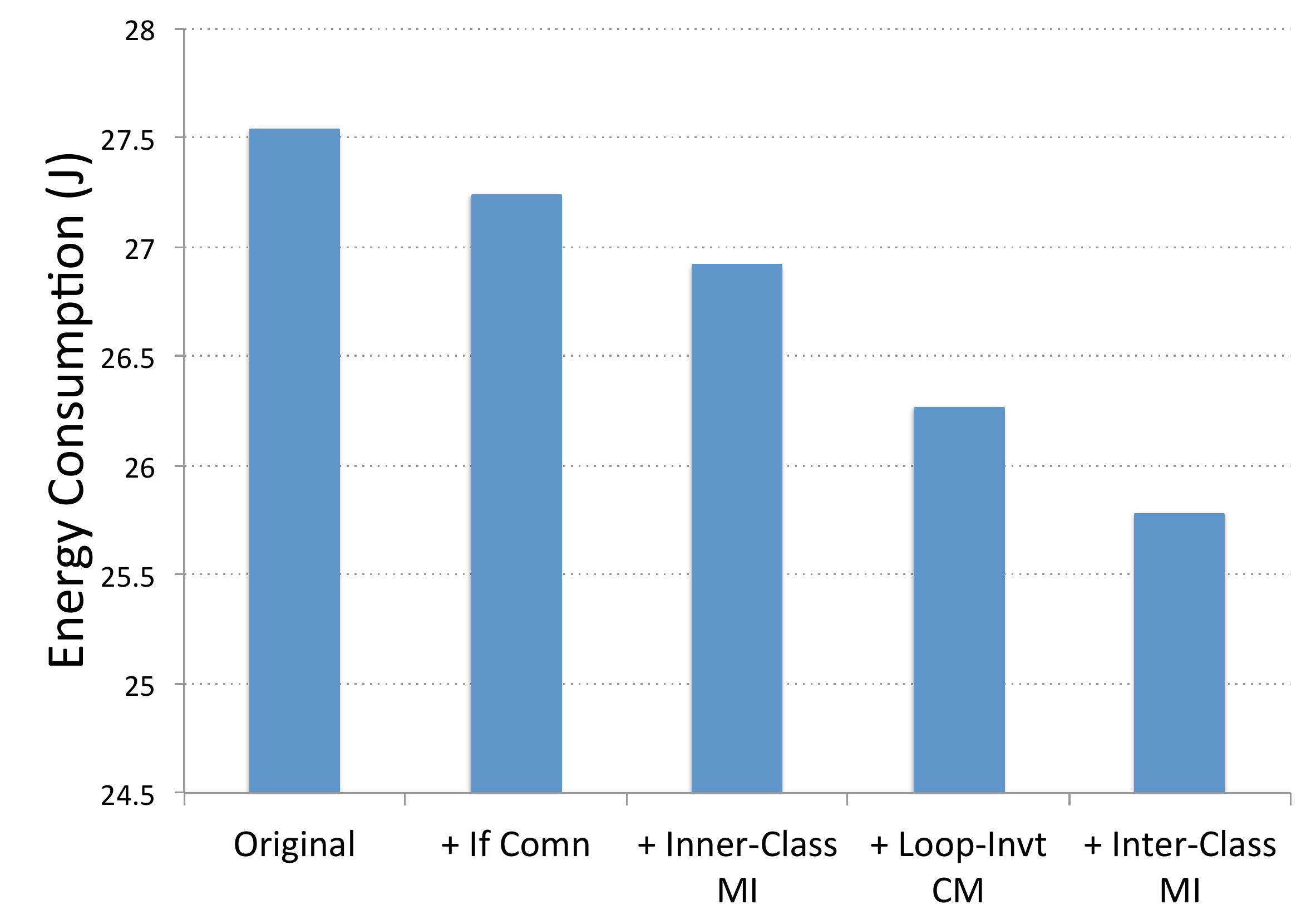}
	\caption{Energy consumption of the code without and with the changes in \texttt{Click \& Move}.}\label{fig:energy_saving_candm}
\end{figure}

Figure \ref{fig:energy_saving_candm} illustrates the energy dissipation of the software without and with the changes introduced in the previous section. 
%And the energy dissipation is measured ten times when the temperature of the CPU is stable at 50$^{\circ}$, we calculate the average value as the result. 
From left to right, the bars indicate cumulative effects of the changes. For example, "\textit{+ If Comn}" is the energy consumption of the original code with the change of "If Combination"; "\textit{+ Inner-Class MI}" is the energy consumption of the code with the changes of both "If Combination" and "Inner-Class Method Inline". 
In total, these four simple changes save 6.4\% of the entire energy consumption without influencing the functionality of code. These changes are made in the basic part of the game engine, which most applications will be bases on, so any gain here can have fundamental impact. Furthermore, these changes are made with little knowledge about the algorithm of the code, the developers who designed the code are surely able to improve the code much more and achieve far more energy saving, if the energy model was available to them.

\section{The \texttt{Orbit} Scenario }\label{Section_orbit}

In this section, we evaluate our approach in the \texttt{Orbit} scenario, where we again refactor the most costly block according to the expensive operations. The experimental result shows that the improvement saves 50.2\% of the CPU energy consumption. 

\subsection{Code Optimization}
\iffalse
\begin{table*}
	\small
	\centering
	\caption{In \texttt{Orbit} scenario, the top 3 costly blocks "In Application". \label{table_lensorbit_topblocks}}
	\begin{tabular}{lrrrrrrrrrr} \hline
		\quad Block ID & \#Executions & Energy Cost (J) & Assi. & Decl. & Cont. & Arra. & Func. & Bool. & Arit. & Libr. \\ 
		\hline%\hline
		CCGrid3D.blit().for\_1 & 330443 \quad\quad &  23839.6 \qquad\quad & 16.7\% & 0\% & 35.6\% & 0\% & 0\% & 20.5\% &  14.0\% & 13.3\% \\
		CCSprite.draw() & 3963 \quad\quad & 385.2 \qquad\quad & 0\% & 0.6\% & 74.6\% & 0\% & 24.8\% & 0\% &  0\% & 0\%\\
		CCNode.visit().if\_4.for\_1 & 8318 \quad\quad& 377.7 \qquad\quad & 24.3\% & 12.9\% & 27.8\% & 0\% & 0\% & 27.3\% &  7.6\% & 0\%\\
		%CCNode.visit() & 7084 \quad\quad& 355.6 \qquad\quad & 0\% & 0\% & 32.1\% & 0\% & 67.9\% & 0\% &  0\% & 0\%\\
		%CCNode.draw()   & 4002 \quad\quad & 197.9 \qquad\quad & 14.6\% & 7.8\% & 30.9\% & 0\% & 46.7\% & 0\% &  0\% & 0\%\\
		\hline
	\end{tabular}
\end{table*}
\fi
\iffalse
\begin{figure}
	\centering
	\includegraphics[width = 0.38\textwidth]{lensorbit_blocks.pdf}
	\caption{The energy proportions of blocks "In Application" in the \texttt{Orbit} scenario}\label{fig:lensorbit_blocks}
\end{figure}
\fi
\subsubsection{Hot Spot Identification}
% As shown in Figure \ref{fig:lensorbit_blocks},

In the \texttt{Orbit} scenario, the block \textit{CCGrid3d.blit().for\_1} dominates the overall energy consumption. 80.9\% of the entire cost is consumed by this block. The second most costly block consumes only 1.3\%. We thus focus attention completely on this single block. 
%"In Application" here means running the \texttt{Orbit} scenario without removing any block.
%It will be seen that a little effort is made to achieve eminent improvements.   

%\subsection{Code Optimization}\label{section:orbit_code_optn}

%Considering \textit{CCGrid3d.blit().for\_1} uses up most of the energy dissipation, so it is the only hot spot we will deal with.

\subsubsection{Strategy Selection \& Code Refactoring}

\begin{program}
	\begin{lstlisting}
	for (int i = 0; i < vertices.limit(); i=i+3) {
    	 mVertexBuffer.put(vertices.get(i)); 
	    mVertexBuffer.put(vertices.get(i+1));
	    mVertexBuffer.put(vertices.get(i+2));
	}
	\end{lstlisting}
	\caption{The \textbf{original} code of \textit{CCGrid3D.blit().for\_1}}
	\label{program7}
\end{program}
\begin{program}
	\begin{lstlisting}
	int limit = vertices.limit(); //added
	for (int i = 0; i < limit; i=i+24) { //changed
	    mVertexBuffer.put(vertices.get(i)); 
	    mVertexBuffer.put(vertices.get(i+1));
	    mVertexBuffer.put(vertices.get(i+2));
	                  ...                       
	    mVertexBuffer.put(vertices.get(i+23));//added
	}
	\end{lstlisting}
	\caption{The changed Program \ref{program7}}
	\label{program8}
\end{program}

Program \ref{program7} shows the original code of \textit{CCGrid3D.blit().for\_1}. In this block, the \textit{Control Ops} (\textit{BlockGoto\_for} and \textit{Field Reference}) use up 35.6\% of the energy; \textit{Boolean Ops} use up 20.5\%; the assignments use up 16.7\%; \textit{Arithmetic Ops} use up 14.0\%; \textit{Lib Functions} use up 13.3\%. We find three easy changes to reduce or replace the pricey operations.

\textit{(i) Loop-Invariant Code Motion:}
In this block, the value of \textit{vertices.limit()} is the constant 2112; we therefore hoist it outside the loop and replace it with the variable \textit{limit}, as shown in Program \ref{program8}. This change avoids invocations and executions of \textit{vertices.limit()} and at the same time decreases a small amount of \textit{Field Reference}. 

\textit{(ii) Loop Unrolling: }
Also as shown in Program \ref{program8}, we duplicate the loop body eight times, reducing the times of comparisons, \textit{BlockGoto\_for}s, assignments and additions. Note that we set the value of the increment as 24 since 24 is a factor of the \textit{limit}, 2112.  

\textit{(iii) Replacement by Library Function: }
The job of Program \ref{program7} or Program \ref{program8} is to get all the elements in \textit{vertices} one by one and put them one by one into \textit{mVertexBuffer}.  Program \ref{program7} can be simply replaced by one line: \textit{mVertexBuffer.put(vertices.asReadOnlyBuffer())}. This puts all the elements of \textit{vertices} into \textit{mVertexBuffer}. This change realizes the same functionality using the existing library function, which is one of the key library functions already compiled into native code. 
\subsection{Evaluation}\label{orbit_evaluation}
\begin{figure}
	\centering
	\includegraphics[width = 0.35\textwidth]{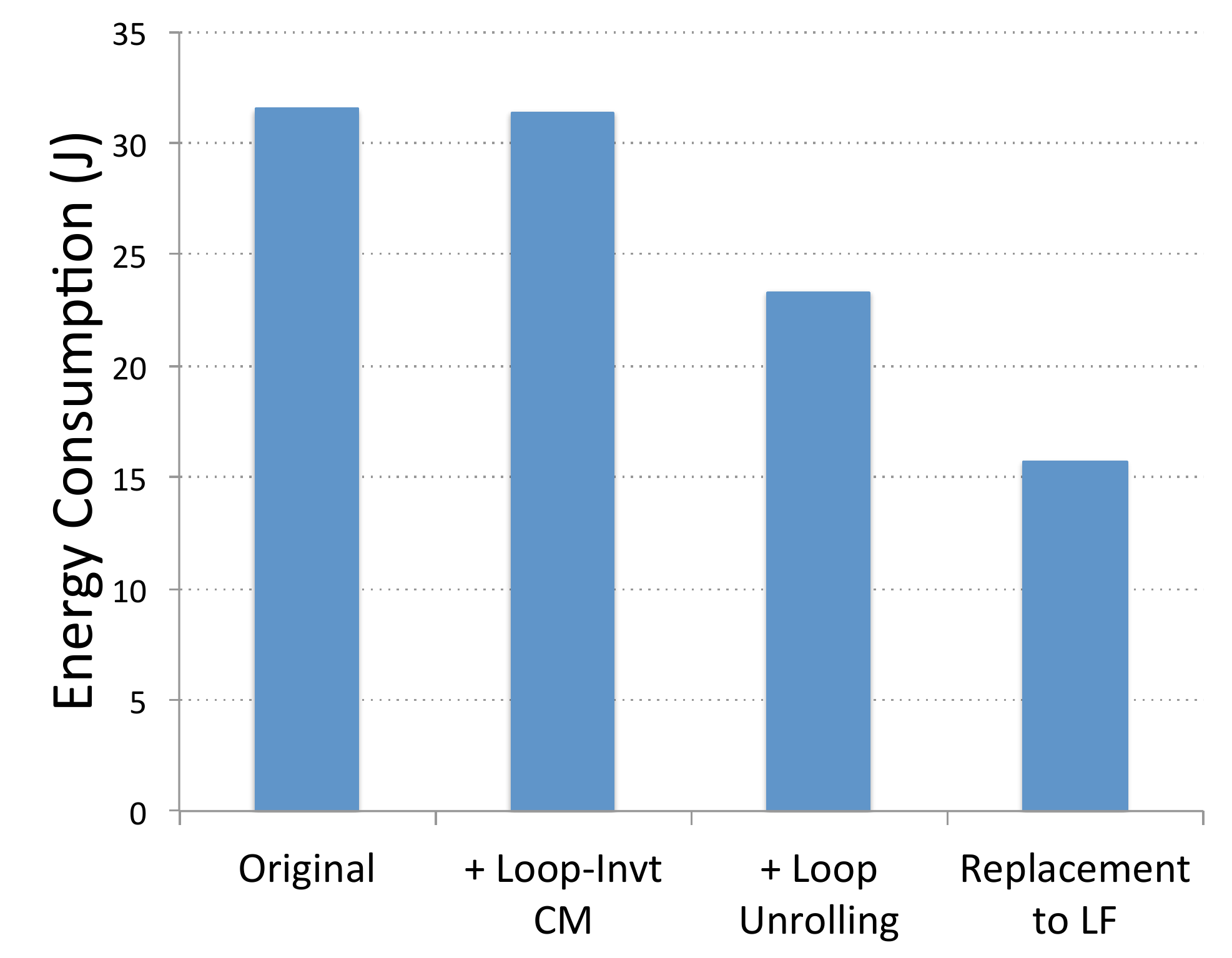}
	\caption{Energy consumption of the code without and with the changes in \texttt{Orbit}.}\label{fig:energy_saving_orbit}
\end{figure}

Figure \ref{fig:energy_saving_orbit} shows the cumulative effects of the code changes on energy consumption. In contrast to the other columns, "\textit{Replacement by LF}" does not take previous changes into account and means only replacing Program \ref{program7} with the built-in library function as stated above. The figure shows that loop-invariant code motion does not gain much energy saving because  \textit{vertices.limit()} is a library function and in addition uses a very small percentage of energy consumption. On the other hand, loop unrolling achieves 25.8\% energy saving due to the reduction of the amount of \textit{Control Ops}, comparisons and assignments, which occupy most of the cost. The most effective change is the replacement by the library function, avoiding the waste of 50.2\% energy use because 
this library function has been compiled into native code before execution; by contrast the Java source code need run-time interpretation which of course incurs an energy cost. The result implies that it is a good idea for developers to make a appropriate use of library functions rather than implementing the same function with Java source code.  %The discovery of this source of inefficiency was assisted by the energy accounting.

\begin{table*}
	\small
	\centering
	\caption{Top 10 most costly blocks "In Application" in the \texttt{Waves} scenario and the energy percentages of different kinds of operations in each block.}\label{table_waves_topblocks}
	\begin{tabular}{lrrrrrrrrr} \hline
		 Block ID & \#Executions & Energy  & Assign & Decl. & Cont. & Func. & Bool. & Arit. & Libr. \\ 
		  & \ & Cost (mJ)&  &  & &  &  &  &  \\ 
		\hline%\hline
		CCGrid3D.blit().for\_1 & 112193  &  8094.1 \quad & 16.7\% & 0\% & 35.6\% &  0\% & 20.5\% &  14.0\% & 13.3\% \\
		CCVertex3D.CCVertex3D() & 40219  & 5232.0 \quad & 27.2\% & 0\% & 10.0\% & 62.8\% & 0\% &  0\% & 0\%\\
		CCWaves3D.update().for\_1.for\_1 & 34604 & 4088.7 \quad & 10.7\% & 0\% & 32.1\% & 0\% & 14.7\% &  39.0\% & 2.2\%\\
		ccGridSize.ccg() & 42275 & 3769.1 \quad & 0\% & 0\% & 32.1\% & 67.9\% & 0\% &  0\% & 0\%\\
		CCGrid3DAction.setVertex()   & 31856  & 3285.4 \quad & 14.6\% & 7.8\% & 30.9\%  & 46.7\% & 0\% &  0\% & 0\%\\
		CCGrid3DAction.originalVertex() & 36566  &  2891.3 \quad & 19.1\% & 10.2\% & 40.3\% & 30.4\% & 0\% &  0\% & 0\%\\
		CCNode.getGrid() & 49119 & 2145.1 \quad & 0\% & 0\% & 58.1\% & 41.9\% & 0\% &  0\% & 0\%\\
		ccGridSize.ccGridSize() & 10570  & 1173.8 \quad & 30.3\% & 0\% & 31.6\% & 38.0\% & 0\% &  0\% & 0\%\\
		CCGrid3D.setVertex() & 3944 & 657.2 \quad & 10.1\% & 1.6\% & 32.8\% & 28.9\% & 0\% &  26.4\% & 0.2\%\\
		CCGrid3D.originalVertex() & 2785 & 374.2 \quad & 14.0\% & 1.9\% & 33.4\% & 17.9\% & 0\% &  32.8\% & 0\%\\ 
		\hline
	\end{tabular}
\end{table*}

\section{The \texttt{Waves} Scenario}\label{Section_waves}

In this section, similarly, we first analyze the energy features of the blocks in the \texttt{Waves} scenario, based on which we modify the code and then evaluate the effects of changes on energy consumption.  

\subsection{Code Optimization}

\subsubsection{Hot Spot Identification}

Unlike the \texttt{Orbit} scenario where only one block dominates energy cost, in the \texttt{Waves} scenario the costs of the top eight blocks are at the same order of magnitude of Joule, as listed in Table \ref{table_waves_topblocks}. The \textit{CCGrid3D.blit().for\_1} is also employed in this scenario and is the most costly as well among all the blocks.

\begin{program}
	\begin{lstlisting}
int i, j;
for( i = 0; i < (gridSize.x+1); i++ ) {
 for( j = 0; j <(gridSize.y+1); j++ ) {
  CCVertex3D v=originalVertex(ccGridSize.ccg(i,j));
             ...
  setVertex(ccGridSize.ccg(i,j), v);
 }
}
	\end{lstlisting}
	\caption{The \textbf{original} code in \textit{CCWaves3D.update()}}
	\label{program9}
\end{program}
\normalsize

\begin{program}
	\begin{lstlisting}
ccGridSize ccgridsize = new ccGridSize(0,0);//added
CCGrid3DAction ccgrid3d = 
       (CCGrid3DAction) target.getGrid(); //added
CCVertex3D	v = new CCVertex3D(0,0,0);     //added
int i, j;
for( i = 0; i < (gridSize.x+1); i++ ) {
 for( j = 0; j <(gridSize.y+1); j++ ) {
  ccgridsize.x=i;ccgridsize.y=j;  //added
  v =ccgrid3d.originalVertex(ccgridsize);//changed
                       ...
  ccgrid3d.setVertex(ccgridsize, v);//changed  
 }
}
	\end{lstlisting}
	\caption{Program \ref{program9} after Method Inline \& Code Motion   }
	\label{program10}
\end{program}

\subsubsection{Strategy Selection \& Code Refactoring}

The majority of blocks in Table \ref{table_waves_topblocks} are directly or indirectly invoked by \textit{CCWaves3D.update().for\_1.for\_1}, as shown in Program \ref{program9}.   The purpose of these methods is mainly to set or get the values of member variables, so a large part of energy consumption goes to assignments, \textit{Function Ops} and \textit{Control Ops}. It was not expected that the code spends such a large amount of energy on simple set and get functions. 

%\subsection{Code Optimization}

\textit{(i) Replacement by Library Function: } We mentioned previously in Section \ref{Section_orbit}  the optimization for \textit{CCGrid3D.blit().for\_1} where we replace the entire Program \ref{program7} with one line of code making use of library functions. We keep this change in this scenario. For other blocks, we come up with a package of modifications as below.

\iffalse
\begin{program}
	\begin{lstlisting}
	...
	for( i = 0; i < (gridSize.x+1); i++ ) {
	ccgridsize.x=i;ccgridsize.y=0;   
	...
	ccgrid3d.setVertex(ccgridsize, v);
	ccgridsize.x=i;ccgridsize.y=1; //added  
	...
	ccgrid3d.setVertex(ccgridsize, v); //added
	...
	...
	ccgridsize.x=i;ccgridsize.y=10; //added  
	...
	ccgrid3d.setVertex(ccgridsize, v); //added         
	}
	\end{lstlisting}
	\caption{Program \ref{program10} after Loop Unrolling }
	\label{program11}
\end{program}
\fi

\textit{(ii) Method Inline \& Code Motion: }
As shown in Program \ref{program9}, the three functions called in the inner loop body are \textit{CCGrid3DAction.originalVertex()}, \textit{ccGridSize.ccg()} and \textit{CCGrid3DAction.setVertex()}, which respectively cost 2891.3 mJ, 3769.1 mJ and 3285.4 mJ when \texttt{Waves} executes with all the blocks. Note that, \textit{CCGrid3DAction} is the parent class of \textit{CCWaves3D}, so in Program \ref{program9} \textit{originalVertext()} and \textit{setVertex()} can be directly called without referring to their class names. 
As seen in Program \ref{program10}, we unpack these three methods in this block: the first and fourth "added" lines are unpacked \textit{ccGridSize.ccg()}; the second "added" and first "changed" lines are unpacked \textit{CCGrid3DAction.originalVertex()}; the second "added" and second "changed" lines are unpacked \textit{CCGrid3DAction.setVertex()}. This change removes all the \textit{Method Invocation}s, parameter passing and value returns related to these three functions invoked by this block. Note that the first three "added" lines are located outside the loop in order to reduce energy consumption of initializing objects and calling \textit{CCNode.getGrid()}. 

%\textbf{Loop Unrolling: }

%As shown in Program \ref{program11}, we multiply the loop body 11 times to remove the inner loop since the value of \textit{gridSize.y} is constantly 10. This modification reduces \textit{BlockGoto\_for} operations, increments and comparisons. 

\subsection{Evaluation}\label{waves_evaluation}

\begin{figure}
	\centering
	\includegraphics[width = 0.35\textwidth]{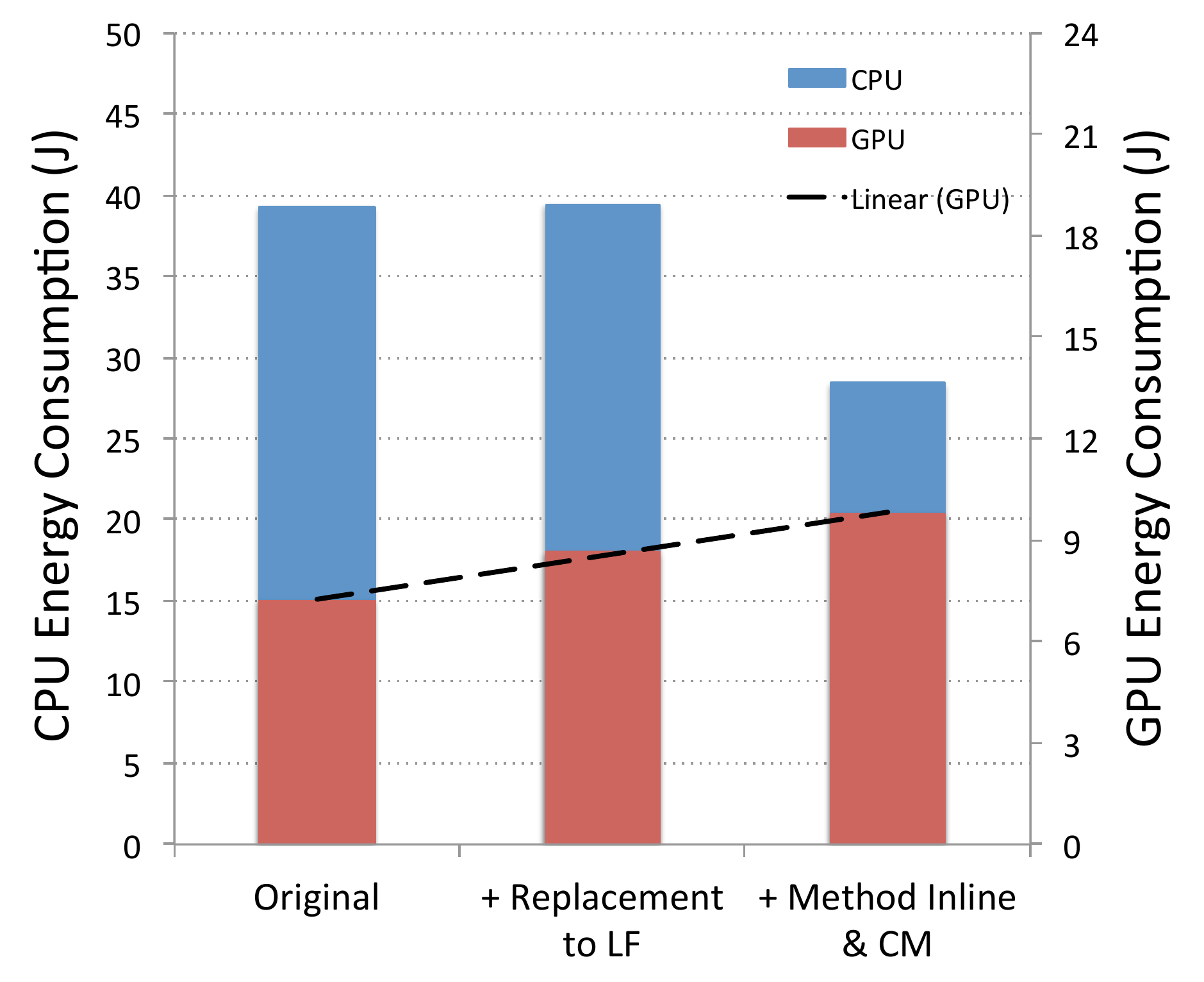}
	\caption{CPU and GPU Energy consumption of the code without and with the changes in \texttt{Waves}.}\label{fig:energy_saving_waves}
\end{figure}

Figure \ref{fig:energy_saving_waves} shows the cumulative effects of changes on energy consumption of CPU and GPU (note that previous figures only showed the CPU energy consumption because the GPU energy consumption did not vary noticeably), and the dashed line indicates the linear trend of the GPU energy consumption.
In the case of games,  the aimed frame rate is usually 60 Hz; when the game overloads the CPU, the rate will decrease, and when the workload is light, even very light, the rate is generally fixed to 60 Hz. The rate in "\textit{Original}" is around 36 Hz; that in "\textit{+ Full-Use LF}" is around 50 Hz; that in "\textit{+ Method Inline \& CM}" is around 60 Hz.
The change of \textit{Full-Use LF} (full use of library function) does not save energy for CPU because the original \texttt{Waves} actually overloads the CPU capacity, so the improvement of code enables the device to generate more frames every second. Consequently, the CPU does the same volume of work and consumes the same amount of energy, while the GPU does more work and consumes more energy, as seen in Figure \ref{fig:energy_saving_waves}. After this change, when we apply the method inline and code motion, 27.7\% of the overall CPU energy is saved, and for the same reason GPU consumes slightly more. This result indicates that our approach not only saves energy but also potentially boosts performance.

%At this moment, the device is capable to fulfill the performance requirement of the improved code, after which when we apply the change of method inline and code motion, 27.7\% of the overall CPU energy is saved. This experimental result shows that our approach not only saves energy but also boosts performance, which benefits users doubly.    

\section{Discussion}

The experiment was targeted successfully at applications using a game engine, which represents one important class of mobile applications, namely games. Further experiment is required to see the extent to which the framework can produce good results on other applications. For instance, in some applications,  energy consumption is distributed more evenly among the blocks, and so the code would not contain such obvious hot spots amenable to optimization. 

Energy modeling is limited to the CPU because the source-level operations identified are CPU-bound. It is another challenge to identify and model source-level operations for GPU, network interface, display and such like. 

Optimization with regard to execution time may result in similar refactorings, if energy use has a strong correlation with execution time as is often the case. However, this observation does not weaken the effectiveness of our approach for saving energy. The novelty of this work is utilizing source-level operation-based information to guide code optimization. The correlation between execution time and energy consumption is beyond the topic of this paper.

%However, this paper aims at designing a energy-optimization framework which is guided by operation-based information. No matter the information is about energy or execution time. In fact, energy information is obviously more relevant. 

\section{Related Work}

A large amount of research effort on energy saving for mobile devices has been focused on the main hardware components, such as the CPU, display and network interface. The CPU-related techniques involve dynamic voltage and frequency scaling \cite{anotherDVFS} and heterogeneous architecture \cite{Reflex_Lin, GreeDroid_Goulding}. Techniques regarding the display include dynamic back-light dimming \cite{dimming_backlight, dimming_backlight2} and tone-mapping based back light scaling \cite{tone_mapping,tone_mapping2}.  Network-related techniques try to exploit idle and deep sleep opportunities \cite{network1,network2}, shape the traffic patterns\cite{network_3,network4}, and etc. Such work attempts to reduce energy dissipation by optimizing the hardware usage at a level below the source code. Besides, several pieces of work aim at designing new hardware and devices \cite{hardwaredisign1, hardwaredesign2}.

On the other hand, there is a significant research engaged in source-level optimization for saving energy. The basic work seeks to understand how the different methods, algorithms and design patterns of software influence the energy consumption.
% The efforts in this part could be related to networking, memory and algorithms. 
For example, \cite{newRoutingTech1,newRoutingTech2,newRoutingTech3} propose new routing techniques and protocols that are aware of energy consumption, which are evaluated by comparing with traditional techniques.
For another example, \cite{choosesortalgorithm} investigates the affects of different sorting algorithms on the energy use with respect to the algorithm's input size. 

Considering design patterns, Litke et al. \cite{desiangPattern1} conduct an experiment showing how big the difference of energy consumption is, before and after the application of design patterns, such as \textit{factory method pattern}, \textit{observer pattern}, and etc. The result reveals that except for one example the use of design patterns does not increase the energy use obviously. Comparable work to \cite{desiangPattern1} is done by \cite{desiangPattern2};  they explore more design patterns and arrive at the conclusion that applying design patterns can both increase and decrease energy consumption, so design-level artifacts cannot be used to estimate the effect of design patterns on energy use.
%the impacts of design patterns on

Vetr\`{o} et al. \cite{code_smell} define the concept of "energy code smells" that are the code patterns (such as self assignment, repeated conditionals and useless control flow) that suggest energy inefficiency. However, the code patterns selected in \cite{code_smell} have little influence (less than 1.0\%) on energy usage.

% Our experimental result shows that our approach is able to save half of the entire energy consumption in certain scenario.   

Regarding code refactoring for energy saving, Ding et al. \cite{energy_saving_programming} perform a small scale evaluation of several commonly suggested programming practices that may reduce energy. Its result shows that reading array length, accessing class field and method invocation all cost remarkable energy. However, this work only provides a small number of tips to developers on how to make the code more energy-efficient. 

Two pieces of work \cite{GeneticImprovement,Seeds} provide systematic approaches to optimizing code. In the former, Boddy et al. attempt to decrease the energy-consumption of software by handling code as if it were genetic material so as to evolve to be more energy-efficient. In the latter, Irene et al. propose a framework to optimize Java application by iteratively searching for more energy-saving implementations in the design space. In summary, these two pieces of work treat the code as a black box, i.e., the optimization is achieved without an analysis on the energy features of the code. In comparison, our approach views the code as a white box since the optimization is guided by the understanding of energy characteristics of source code.

Apart from the energy modeling, our instantiation of the framework is literally profile-guided, however, employs the Operation-Based information (the standard profile technique \cite{Simunic:2000:source_code_optimization} can not access) to connect the understanding of code features with very Targeted refactoring strategy for such a high-level source code as Java. By comparison, the standard profile-based technique can only tell the hot spots at a size as smallest as methods or functions and hardly indicate the very targeted solution. 
%no connection between understanding and optimization
%to the best of our knowledge 
%The experimental evaluation demonstrates that the approach is an effective and practical approach to energy-aware mobile application development.

%Moreover, our experiment is conducted on real application rather than 

%1) we construct the operation-based source-code-level energy model; 2) based on the model, we capture the energy characteristics of the code; 3) we improve the code by removing, reducing or replacing the expensive operations in the costly blocks.    
%correlating the operations to the energy cost by case analysis. Thus our model does not require  the profile of target system and more flexible to various pieces and types of code. The model is also capable to help produce the energy breakdown of the code to direct the developer's effort on energy efficiency.      

\section{Conclusion}

%In this paper, we propose an energy-aware programming approach for mobile app development, guided by an operation-based source-level energy model. The approach consists of 1)  construction of an operation-based energy model by mining the data generated in a range of well-designed execution cases; 2) capturing energy characteristics of the code based on the model; 3) improving the code by removing, reducing or replacing the expensive operations in the costly blocks.   
This paper presents a source-level energy-optimization framework, guided by the understanding of energy features of source code. The framework constructs the infrastructure for code optimization by both automatic tools and developers. We also implement an instantiation of the framework driven by a source-level operation-based energy model. 

We evaluate the instantiation on a physical Android development board with two ARM quad-core CPUs and on a real-world game engine. In the case study our approach has a significantly impact on energy saving. In different scenarios, this approach saves CPU energy consumption up to 50.2\%. %The findings also indicate that the performance of code is a potential by-product of this approach, which improves the user experience more.\\\

%fine-grained energy model for mobile application source code on the basis of energy operations. We first introduce the energy operations that are identified directly from the source code. The energy operations are employed as the basic units that constitute the overall energy consumption of the source code. We then design a wide diversity of execution cases to generate data about the operation executions and the entire energy consumption. Regression analysis is applied to use the data to estimate the energy consumption of each operation. Finally, we show that the model is capable to capture comprehensive energy features that coarse-grained models or techniques could not shed light on.
%\section{Acknowledgements}

%This research is funded by the European Union Seventh Framework Programme (FP7/2007-2013) under grant agreement no 318337, ENTRA - Whole-Systems Energy Transparency.


\bibliographystyle{abbrv}
\bibliography{sigproc}  % sigproc.bib is the name of the Bibliography in this case

\end{document}